\documentclass[aps,prb,twocolumn,superscriptaddress,longbibliography]{revtex4-1}
\usepackage[dvipsnames]{xcolor}
\usepackage{graphicx}
\usepackage{amsmath,amssymb}
\usepackage{bm}
\usepackage{braket}
\usepackage{lipsum}
\usepackage{makerobust}
\usepackage{simpler-wick}
\usepackage{comment}
\usepackage{multirow}
\usepackage{dcolumn}
\usepackage{calc}

\newcolumntype{L}{D{.}{.}{2,6}}
\newcommand{\spbox}[2]{\makebox[\widthof{#1}]{#2}}

\newcommand*\circled[1]{\tikz[baseline=(char.base)]{
    \node[shape=circle,draw,inner sep=1pt] (char) {#1};}}
\MakeRobustCommand\circled

\newcommand{\makeauthor}[2]{\newcommand{#1}[1]{{%
  \sffamily\color{#2}{%
    \bfseries\begingroup\escapechar=-1\edef\x{\endgroup\string#1}\x:%
  } ##1}}%
  \MakeRobustCommand#1}
\makeauthor{\eric}{Plum}
\makeauthor{\themba}{ForestGreen} 
\makeauthor{\sr}{blue}
\makeauthor{\Fig}{red}

\begin{document}

\renewcommand{\vec}[1]{\bm{#1}}
\newcommand{\up}{{\uparrow}}
\newcommand{\dw}{{\downarrow}}
\newcommand{\pa}{{\partial}}
\newcommand{\pd}{{\phantom{\dagger}}}
\newcommand{\bs}[1]{\boldsymbol{#1}}
\newcommand{\add}[1]{{{\color{black}#1}}}
\newcommand{\edit}[1]{{{\color{magenta}#1}}}
\newcommand{\lucca}[1]{{{\color{orange}#1}}}
\newcommand{\mb}[1]{{{\color{violet}#1}}}
\newcommand{\todo}[1]{{\textbf{\color{red}ToDo: #1}}}
\newcommand{\tbr}[1]{{\textbf{\color{red}\underline{ToBeRemoved:} #1}}}
\newcommand{\eps}{{\varepsilon}}
\newcommand{\nn}{\nonumber}
\def\ie{\emph{i.e.},\ }
\def\eg{\emph{e.g.},\ }
\def\ea{\emph{et. al.}\ }
\def\cf{\emph{cf.}\ }

\newcommand{\brap}[1]{{\bra{#1}_{\rm phys}}}
\newcommand{\bral}[1]{{\bra{#1}_{\rm log}}}
\newcommand{\ketp}[1]{{\ket{#1}_{\rm phys}}}
\newcommand{\ketl}[1]{{\ket{#1}_{\rm log}}}
\newcommand{\braketp}[1]{{\braket{#1}_{\rm phys}}}
\newcommand{\braketl}[1]{{\braket{#1}_{\rm log}}}

\graphicspath{{./}{./figures/}}



\title{Electronic structure, spin-orbit interaction and electron--phonon coupling\\
of triangular adatom lattices on semiconductor substrates}

\author{Lucca Marchetti}
\affiliation{School of Physics, University of Melbourne, Parkville, VIC 3010, Australia}
\author{Matthew Bunney}
\affiliation{School of Physics, University of Melbourne, Parkville, VIC 3010, Australia}
\affiliation{Institute for Theoretical Solid State Physics, RWTH Aachen University, 52062 Aachen, Germany}
\author{Domenico Di Sante}
\affiliation{Department of Physics and Astronomy, University of Bologna, 40127 Bologna, Italy}
\author{Stephan Rachel}
\affiliation{School of Physics, University of Melbourne, Parkville, VIC 3010, Australia}

\date{\today}

\begin{abstract}
A one-third monolayer of the heavy metals Sn and Pb deposited on semiconductor substrates can lead to a $\sqrt{3}\times\sqrt{3}$ surface reconstruction, constituting an exciting triangular lattice material platform.
A long history of experiments identified charge-ordered and magnetic ground states. These discoveries were accompanied by a decades-long debate of whether electron correlations  or other effects involving phonons are the driving force of the symmetry-broken states. The most recent discovery of superconductivity in boron-doped Sn/Si(111) with a $T_c$ between 5K and 9K led to a renewed excitement. Here we revisit the electronic and phononic properties of Sn and Pb adatom triangular lattices on Si(111) 
and SiC(0001). For all materials we compute relativistic bandstructures using DFT+$U$ where $U$ is only applied to the substrate atoms in order to adjust the band gap to match the experimental value; as a consequence, some of the resulting tight-binding parameters of the metallic surface band differ substantially compared to previous studies. Remarkably, for Pb/SiC(0001) we predict Rashba spin-orbit coupling as large as 45\% of the nearest-neighbor hopping energy. In addition, we compute the phonon spectra and electron-phonon coupling constants for all materials, and for Pb/Si(111) even relativistically although the inclusion of spin-orbit coupling has surprisingly little effect on the electron-phonon coupling constant. We conclude that the resulting couplings are too weak to account for electron-phonon mediated superconductivity in any of these materials.
\end{abstract}

\maketitle

%
%

\section{Introduction}\label{sec:intro}


The material class of a 1/3 monolayer of group-IV adsorbates (Pb, Sn) on semiconductor substrates has a long history. The 1/3 monolayer coverage leads to a  $\sqrt{3}\times\sqrt{3}R30^\circ$ reconstruction where the atoms sit on the $T_4$ sites in a triangular lattice. Thus three out of their four valence bonds are saturated with the substrate, while the fourth orbital remains ``dangling'' and leads to a half-filled surface band, susceptible to electron-electron interactions. The
 symmetry-broken low-temperature ground states have fascinated researchers for decades. It started with Pb/Ge(111) and the observation of a low-temperature surface charge density wave\,\cite{carpinelli-96n398}. The same group of researchers found a similar phenomenology for the sister compound Sn/Ge(111)\,\cite{carpinelli-97prl2859}. However, the transition from the room temperature $\sqrt{3}\times\sqrt{3}$ phase to the low temperature $3\times 3$ phase associated with charge-ordering could not be explained using first-principles methods, in contrast to Pb/Ge(111)\,\cite{carpinelli-96n398}. The authors conjectured\,\cite{carpinelli-97prl2859} that electron correlations might be responsible for the charge ordering in Sn/Ge(111). Also the role of defects was discussed in this context\,\cite{weitering-99s2107,melechko-00prb2235}.

The importance of electron-electron interactions was discussed from the beginning\,\cite{carpinelli-97prl2859,santoro-99prb1891}, but probably the first paper to propose and explicate a Mott--Hubbard scenario was Ref.\,\onlinecite{profeta-07prl086401}. By performing local density and Hubbard-$U$ approximation, they proposed both Sn/Ge(111) and Sn/Si(111) as Mott insulating materials. \add{In Ref.\,\onlinecite{schuwalow-10prb035116} electronic correlations in Sn/Ge(111) and Sn/Si(111) were considered beyond perturbative approximations within dynamical mean-field theory.}
Ref.\,\onlinecite{cortes-13prb125113} also reported the charge-ordered $3\times3$ phase in Sn/Ge(111), and characterized it as an intermediate phase between the metallic $3\times3$ and Mott insulator phases.
The importance of spin-orbit coupling (SOC) was emphasized in Ref.\,\onlinecite{tresca-21prb045126} for Pb/Ge(111) and Pb/Si(111).


Sn/Si(111) was originally believed to feature surface-metallicity, but micro-four-point-probe conductivity measurements revealed insulating behavior from room temperature all the way down to low temperatures\,\cite{hirahara-09prb235419,xiongExperimentalObservationPseudogap2022}. Doping by partially replacing Sn atoms with In or Na deposition, leads rather suddenly to a metallic conductivity suggesting a small energy gap of the undoped Sn/Si(111) system. Angle-resolved photoemission spectroscopy  performed on Sn/Si(111) revealed that the measured spectral function is incompatible with both $3\times 3$ and $\sqrt{3}\times\sqrt{3}$ reconstruction, but compatible with a $2\sqrt{3}\times 2\sqrt{3}$ phase associated with a row-wise antiferromagnetic order\,\cite{li-13nc1620}. The accompanying many-body calculations estimated a quite sizeable Hubbard-$U=0.66$eV.
The insulating ground state of Sn/Si(111) was also suggested as a consequence of a Slater-type insulator via band magnetism\,\cite{lee-14prb125439}. Ref.\,\onlinecite{hansmann-13prl166401} performed a fully first-principles approach, predicting the size of local and non-local Coulomb interactions and derived phase diagrams for various Y/Si(111) compounds (Y=Si, C, Sn, Pb). For Sn/Si(111) they found local Hubbard-$U=1$eV and nearest-neighbor $V_1=0.5$eV compared to a bandwidth $W\sim 0.5$eV, further establishing Sn/Si(111) as a Mott insulator.
The importance of SOC for Y/Si(111) was emphasized in Ref.\,\onlinecite{badrtdinov-16prb224418}, where  SOC combined with strong electron correlations was even proposed to stabilize magnetic skyrmions due to Dzyaloshinskii-Moriya interactions.

The observation of $3\times 3$ to $\sqrt{3}\times\sqrt{3}$ reversible phase transition in Pb/Si(111) by means of variable temperature scanning tunneling microscopy (STM) and DFT calculations was reported in Ref.\,\cite{brihuega-05prl046101,tresca-18prl196402,adler-19prl086401,horikoshiStructuralPhaseTransitions1999}
; the phase transition was found at $T_c=86$K.
STM and quasi-particle interference (QPI) experiments combined with DFT$+U$ simulations for Pb/Si(111) reported a chiral texture in the charge-ordered phase\,\cite{tresca-18prl196402}. Phonons along with the strong SOC were proposed as a driving mechanism for the observed charge order. In contrast, a combined STM and QPI study with many-body variational cluster approach simulations emphasized the relevance of non-local Coulomb interactions for the charge-order in Pb/Si(111)\,\cite{adler-19prl086401}.

\begin{figure*}
    \centering
    \includegraphics{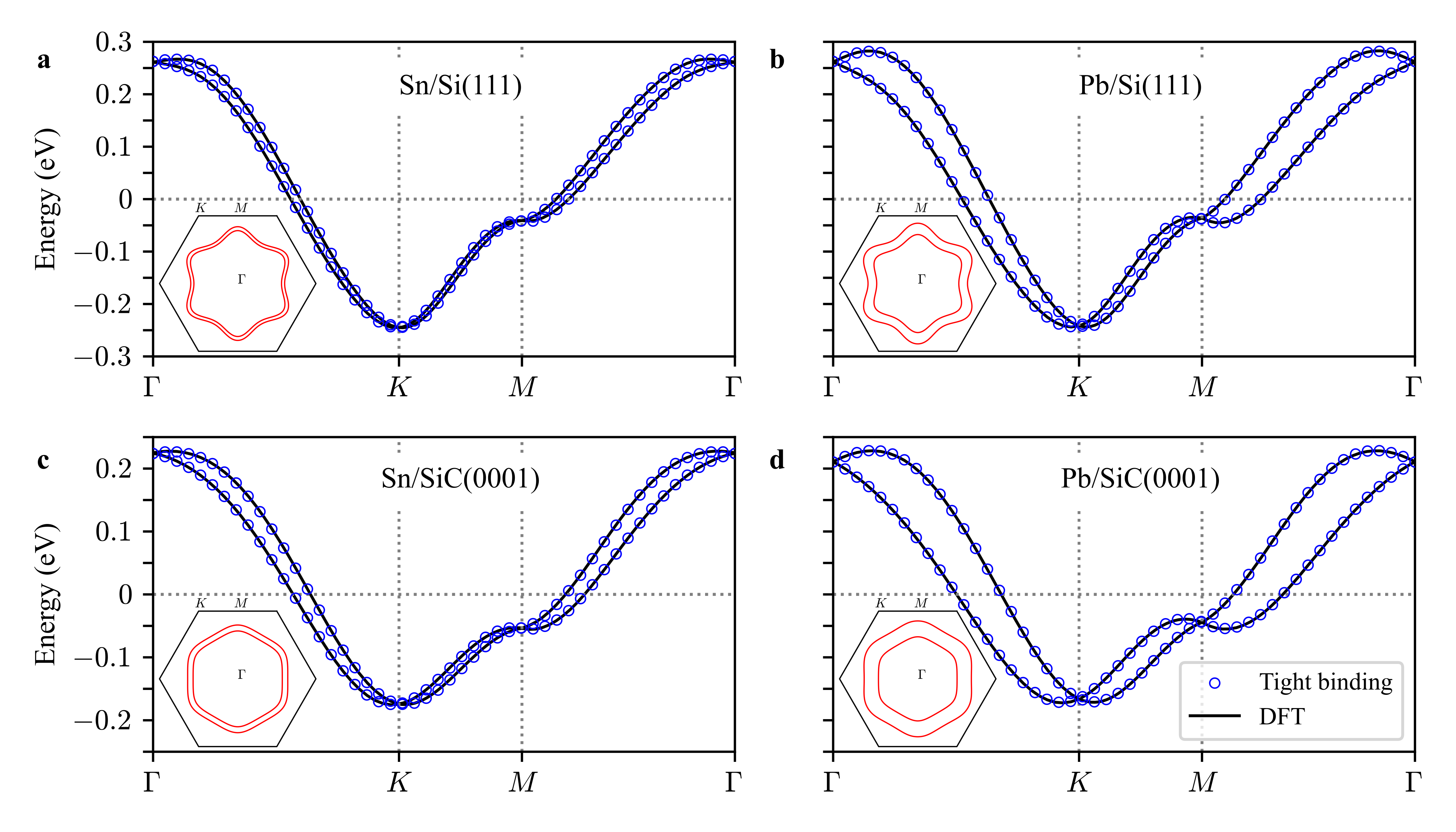}
    \caption{
    DFT band structure and extracted tight binding model for \textbf{a} Sn/Si(111), \textbf{b} Pb/Si(111), \textbf{c} Sn/SiC(0001) and \textbf{d} Pb/SiC(0001). DFT+$U$ has been used to match experimental band gaps for the substrates. Tight binding models have been constructed using the hopping parameters listed in Tab.\,\ref{table:Hopping parameters, with  DFT+U SOC} and SOC splitting in Tab.\,\ref{table:Rashba splitting DFT+U}, \ref{table:Dot product splitting DFT+U} and \ref{table:DMI splitting DFT+U}. \add{Fermi surfaces at half-filling calculated from the tight-binding models are included as insets.}}
    \label{fig:DFT+U_TB}
\end{figure*}

Successful hole-doping of Sn/Si(111)\,\cite{ming-17prl266802} and indications for superconductivity\,\cite{ming-18prb075403} were reported, supporting the scenario of a doped Mott insulator. 
\add{At the same time, theoretical work proposed chiral $d$-wave superconductivity in Sn/Si(111)\,\cite{cao-18prb155145}.}
Shortly after, the discovery of superconductivity with $T_c=4.7$K in boron-doped Sn/Si(111) marks a breakthrough\,\cite{wu-20prl117001}.
Follow-up theory work suggested that the superconducting ground state might be unconventional and constitute chiral topological superconductivity\,\cite{wolf-22prl167002,biderang-22prb054514}, its nature depending on filling, local and non-local Hubbard interactions. Most recently, experiments on B-doped Sn/Si(111) reported the realization of chiral superconductivity\,\cite{ming-23np500} with critical temperatures 
close to $T_c \sim 10$K.


Adatom lattices on SiC(0001) are the least studied materials amongst our candidates. Previous DFT calculations of Sn/SiC(0001) closely reproduce STM measurements\,\cite{glass-15prl247602}. Photoemission data show a deeply gapped state with a gap size of $\sim 2$eV; based on dynamical mean-field theory it is argued that it reflects a pronounced Mott-insulating ground state, possibly with antiferromagnetic ordering.

\begin{table}[b]
\centering
\begin{tabular}{c L L L L}
 \hline\hline
  \multicolumn{1}{c}{Hopping} & \multicolumn{1}{c}{Sn/Si(111)} & \multicolumn{1}{c}{Pb/Si(111)} & \multicolumn{1}{c}{Sn/SiC(0001)} & \multicolumn{1}{c}{Pb/SiC(0001)} \\
 \hline
    $t_0$ (eV) & 2.229660 & 2.211626  & 5.751530 & 5.739806 \\

    $t_1$ (eV)& 0.052773 & 0.052115  & 0.043110 & 0.040175 \\

    $t_2/t_1$ & -0.270290 & -0.276388  & -0.198260 & -0.204207 \\

    $t_3/t_1$ & 0.097436 & 0.100489  & 0.019949 & 0.040124 \\

    $t_4/t_1$ &  -0.018305 & -0.017577  & 0.002644 & -0.000548 \\

    $t_5/t_1$ & 0.019442 & 0.020532  & -0.003155 & -0.004828 \\

    $t_6/t_1$ & -0.030868 & -0.031987  & -0.003897 & -0.004107 \\

    $t_7/t_1$ & 0.006386 & 0.007503  &  & \\

    $t_8/t_1$ & -0.002312 & -0.002437  &  & \\

    $t_9/t_1$ & -0.002350 & -0.002322  &  & \\
\hline\hline

\end{tabular}
 \caption{Tight binding hopping parameters from relativistic DFT+$U$ calculations as described in the main text. Calculations for X/Si(111) required a $10\times10$ kpoint grid for good fitting quality and hence has longer range hoppings included compared to X/SiC(0001) which used a  $7\times7$ kpoint grid.}
 \label{table:Hopping parameters, with  DFT+U SOC}
\end{table}

The aim of this paper is to revisit the band structures of the materials Sn/Si(111), Pb/Si(111) and Sn/SiC(0001) in their $\sqrt{3}\times\sqrt{3}$-reconstructued surface phase with a focus on the strength of spin-orbit coupling (SOC) in Sec.\,\ref{sec:results1}; we complement our investigation with Pb/SiC(0001) whose $\sqrt{3}\times\sqrt{3}$-phase has not been reported so far. We do not investigate many-body ground states involving magnetic or charge order using DFT, but rather study and analyze the non-interacting metallic bandstructures. 
We further analyze phonon spectra of these materials in Sec.\,\ref{sec:results2} in order to evaluate the electron-phonon coupling (EPC) strength $\lambda$ in Sec.\,\ref{sec:results3} A relativistic study of EPC for Pb/Si(111) is presented in Sec.\,\ref{sec:results4}.
The discussion of our results is presented in Sec.\,\ref{sec:discussion}, before the end with an outlook in Sec.\,\ref{sec:outlook}.
\add{In Appendix \ref{sec:App-Ge} we present analogous results of the electronic and phononic properties as well as EPC for Sn/Ge(111) and Pb/Ge(111).}

%
%

%
%

\section{Results I: Electronic structure}\label{sec:results1}
The electronic band structure of the four materials with the exception of Pb/SiC(0001) has been reported previously\,\cite{tresca-18prl196402,glass-15prl247602,badrtdinov-16prb224418,badrtdinovNanoskyrmionEngineeringElectron2018,hansmann-13prl166401,adler-19prl086401}. 
Here we revisit and compare the four materials. We are interested in their metallic, non-interacting surface bandstructures. By deriving precise tight-binding models of these bands, we provide valuable input for any future many-body analysis of magnetic, charge-ordered, superconducting or otherwise ordered ground states.
While the DFT bandstructures of the four materials are rather similar, it turns out that they realize different regimes of SOC ranging from 10\% to 45\% of the nearest-neighbor hopping amplitude $t_1$. This might have major implications for the (strongly) correlated many-body ground states of these materials, as pointed out in a recent study for the plain-vanilla triangular-lattice Rashba-Hubbard model\,\cite{bunney-PhysRevB.110.L161103}.

\begin{table}[b]
\centering
\begin{tabular}{c c c c c}
 \hline\hline
   SOC & Sn/Si(111) & Pb/Si(111) & Sn/SiC(0001) & Pb/SiC(0001) \\
 \hline
    $\alpha_1 / t_1$ & 0.102860 & 0.294286  & 0.166067 & 0.444003 \\

    $\alpha_2 / t_1$ & 0.006051 & 0.014123  & 0.012804 & 0.031794 \\

    $\alpha_3 / t_1$ & 0.002577 & 0.002744  & 0.003062 & 0.013889 \\

    $\alpha_4 / t_1$ & 0.000697 & 0.002570  & 0.000379 & 0.000707 \\

    $\alpha_5 / t_1$ &  0.000733 & 0.001983  & 0.000077 & 0.000133 \\

    $\alpha_6 / t_1$ & 0.000405 & 0.001130  & 0.000174 & 0.000438 \\

    $\alpha_7 / t_1$ & 0.000184 & 0.000348  &  &  \\
    $\alpha_8 / t_1$ & 0.000022 & 0.000004  &  &  \\
    $\alpha_9 / t_1$ & 0.000035 & 0.000169  &  &  \\
\hline\hline

\end{tabular}
 \caption{Rashba SOC parameters using DFT+$U$. Parameters are given as ratio of $\alpha_n / t_1$ for $n$th nearest neighbor splitting.}
 \label{table:Rashba splitting DFT+U}
\end{table}

We use the ab initio package VASP\,\cite{kresseEfficientIterativeSchemes1996} within the Perdew-Burke-Ernzerhof (PBE) \textit{generalized gradient approximation} (GGA) \cite{perdewGeneralizedGradientApproximation1996} to compute the relativistic electronic bandstructures of the four materials. In particular, we employ DFT$+U$ using the approach of Dudarev \ea \cite{dudarevElectronenergylossSpectraStructural1998} where the Coulomb potential $U$ is applied only to the Si or SiC substrate atoms, but not on the Sn and Pb adatoms. 
The main effect of $U$ is to vary the gap size of the semiconductor substrate; only through changes of the gap size is the metallic surface band influenced. \add{However, the surface band remains sufficiently uncorrelated within our first-principles calculations.}
The values for $U$ to get the correct band gaps are $U_{\rm Sn/Si(111)}=-2.84$\,eV,  $U_{\rm Pb/Si(111)}=-3.17$\,eV,  $U_{\rm Sn/SiC(0001)}=-11.64$\,eV and $U_{\rm Pb/SiC(0001)}=-11.64$\,eV.

\add{
DFT, particularly in its local density and gradient-corrected local density approximations, is known to inadequately describe the exact exchange-correlation hole. 
The exchange-correlation hole represents the probability distribution of finding an electron at position $r_1$ given the presence of another electron at position $r_2$.
It is intuitive to consider that electronic correlations significantly influence the shape and profile of the exchange-correlation hole.
By comparing numerically exact exchange-correlation holes with those obtained via DFT in simple systems, it becomes evident that the local density approximation underestimates the hole in regions near the atomic core, where states are more localized.
To satisfy the probability sum rule, this underestimation results in a corresponding overestimation at greater distances, in regions dominated by delocalized states\,\cite{gunnarssonDescriptionsExchangeCorrelation1979}.
For strongly localized states, the DFT+$U$ method (with positive $U$) enhances the exchange-correlation hole by maintaining electrons at greater reciprocal distances.
Conversely, in delocalized states, a correction can be achieved with a negative $U$ value, which effectively reduces electron-electron interactions.
In small band-gap semiconductors like Si and SiC, where the band gap arises from delocalized $p$-like electronic states, reducing the effective Coulomb repulsion (via a negative $U$) increases the overlap of electron wavefunctions.
This, in turn, modifies the bonding-antibonding interactions and adjusts the associated band gap.
Alternative exchange-correlation functionals, such as hybrid functionals\,\cite{HeydHybridFunctionals2003,AdamoReliableDensity1999}, address the local density approximation's shortcomings by incorporating a fraction of exact exchange, which mitigates the self-interaction problem.
However, the computational cost of exact exchange is prohibitive, especially for systems comprising 40 atoms with dense $k$-point sampling, such as adatoms on Si(111) or SiC(0001).
Meta-GGA functionals\,\cite{TaoClimbingDensity2003,PerdewWorkhorse2009,BeckeSimplePotential2006,TranAccurateBandGaps2009} also improve band-gap estimates by incorporating the kinetic energy density into the exchange-correlation potential.
Nonetheless, their application to large supercell models involving surfaces and vacuum regions is not yet standard practice. These aspects led us to opt for the accurate and widely adopted DFT+$U$ approach.
}

In these calculations we have used a plane-wave cutoff energy of 600\,eV and energy convergence criteria of $10^{-8}$eV. 
For X/SiC(111) $7\times7\times1$ $\Gamma$-centered grids were used to sample the Brillouin zone, whereas for X/Si(111) $10\times10\times1$ $\Gamma$-centered grids were used for better tight-binding fits close to the $\Gamma$-point.
\add{
Atomic positions within the cell were based off experimental values and then optimized within VASP. 
A force criteria of 0.0001 \AA/eV was used in the relaxation procedures.
For the X/SiC(0001) materials, the cell volume was also optimized to relieve excess pressure on the cell.
In the relaxed cells, the adatoms sit above the bulk at heights of $h = 2.48 \,{\rm \AA}$ for Sn/Si(111), $h = 2.62 \,{\rm \AA}$ for Pb/Si(111), $h = 2.10 \,{\rm \AA}$ for Sn/SiC(0001), and $h = 2.21 \,{\rm \AA}$ for Pb/SiC(0001).
Approximately 9.5\,{$\rm \AA$} and 2.6\,{$\rm \AA$} of vacuum were used for the X/Si(111) and X/SiC(0001) materials, respectively.
}
The DFT$+U$ bandstructures along with the extracted tight-binding bands (discussed in detail below) are shown in Fig.\,\ref{fig:DFT+U_TB}, zoomed in on the energy range of the metallic surface band. We find the bandwidths
$W/t_1 = 9.71$ ($W=0.51$eV) for Sn/Si(111),
$W/t_1 = 7.69$ ($W=0.53$eV) for Pb/Si(111),
$W/t_1 = 9.29$ ($W=0.40$eV) for Sn/SiC(0001) and
$W/t_1 = 9.97$ ($W=0.40$eV) for Pb/SiC(0001).
In Appendix~\ref{sec:app-electronicstructure} we show in Fig.\,\ref{fig:SM-bandstructure} the same bandstructure plots with larger energy range involving substrate bands.
\add{
The resulting Fermi surfaces are included as insets in Fig.\,\ref{fig:DFT+U_TB}.
The Fermi surfaces of the X/Si(111) materials take a more warped hexagonal shape compared with the X/SiC(0001) materials due to the larger further nearest neighbor hopping values.
}

\begin{table}[t!]
\centering
\begin{tabular}{c c c c c}
 \hline\hline
   SOC & Sn/Si(111) & Pb/Si(111) & Sn/SiC(0001) & Pb/SiC(0001) \\
 \hline
    $\beta_4 / t_1$ & 0.001133 & 0.003756  & 0.000211 &0.001051 \\

    $\beta_7 / t_1$ & $-0.000089$ & $-0.000223$  &  &  \\

    $\beta_9 / t_1$ & 0.000068 & 0.000225  &  &  \\
\hline\hline

\end{tabular}
 \caption{Radial Rashba SOC parameters present for 4th-, 7th-, 9th-nearest neighbor shells.}
 \label{table:Dot product splitting DFT+U}
\end{table}

The four considered materials constitute a case where the metallic surface band is energetically completely separated from the substrate bands, and the description as a single-band tight-binding model is fully justified. The corresponding tight-binding parameters are listed in Tab.\,\ref{table:Hopping parameters, with  DFT+U SOC} and in Tab.\,\ref{table:Rashba splitting DFT+U}.
\add{
The tight-binding parameters are obtained by projecting Wannier functions onto the DFT bandstructures.
Maximally localized Wannier functions are obtained using the Wannier90 software\cite{pizziWannier90CommunityCode2020}, initially projecting onto the adatom $p_z$ orbitals.
}

While the isolated triangular lattice of surface adatoms carries a $D_{6h}$ point group symmetry, the presence of the Si(111) or SiC(0001) substrate breaks inversion symmetry, and reduces the rotational symmetry from $C_6$ to $C_3$. This reduces the point group symmetry to $C_{3v}$. Group theoretic techniques were used to decompose the Hamiltonian terms, implementational details of which we delegate to Appendix~\ref{sec:app-electronicstructure}. Further symmetry analysis of the model has consequences discussed further below.
 Interestingly, we find for the nearest-neighbor Rashba amplitude $\alpha_1/t_1$ the value 0.10 for Sn/Si(111), 0.17 for Sn/SiC(0001), 0.29 for Pb/Si(111) and 0.44 for Pb/SiC(0001). The SOC splitting in Sn/SiC(0001) was previously reported to be $13.7\%$\,\cite{glass-15prl247602}. We attribute the increase in SOC strength in the X/SiC(0001) materials compared to their X/Si(111) counterparts due to the lack of inversion symmetry in the bulk SiC compared to bulk Si. The tight-binding Hamiltonian for our systems 
 can be written as

 \begin{table}[t!]
\centering
\begin{tabular}{c c c c c}
 \hline\hline
   SOC & Sn/Si(111) & Pb/Si(111) & Sn/SiC(0001) & Pb/SiC(0001) \\
 \hline
    $J_2 / t_1$ & $-0.002596$ & $-0.002590$  & $-0.003363$ & $-0.007169$ \\

    $J_4 / t_1$ & 0.000341 & 0.001708  & 0.000336 & 0.001419 \\

    $J_6 / t_1$ & $-0.000227$ & $-0.000154$  & $-0.000046$ & $-0.000025$ \\

    $J_7 / t_1$ & $-0.000171$ & $-0.000652$  &  &  \\

    $J_9 / t_1$ & 0.000000  & 0.000134  &  &  \\
\hline\hline

\end{tabular}
 \caption{Valley Zeeman parameters present for nearest neighbor shells with sites lying off the $C_{3v}$ mirror planes determined by the bulk substrate.}
 \label{table:DMI splitting DFT+U}
\end{table}
\begin{equation}
\label{eqn:ham}
    H = H_0 + H_R + H_{\rm{VZ}}
\end{equation}
with
\begin{equation}
    H_0 = \sum\limits_{i,j} \sum\limits_{\sigma} t_{ij} c^\dagger_{i\sigma} c^\pd_{j\sigma}
\end{equation}
\vspace{-4mm}
\begin{equation}
    H_R = \sum\limits_{i,j} \sum\limits_{\sigma,\sigma'} i\lambda_{ij} c^\dagger_{i\sigma} c^\pd_{j\sigma'} [\Upsilon_{ij}^\ast
    (\bm{\sigma}\times\bm{r}_{ij})\cdot\hat{\bm{z}}\,\Upsilon_{ij}]_{\sigma\sigma'}
\end{equation}
\vspace{-4mm}
\begin{equation}
\label{eqn:hvz}
    H_{\rm{VZ}} = \sum\limits_{i,j} \sum\limits_{\sigma,\sigma'} i\nu_{ij}J_{ij}{\sigma^z}_{\sigma\sigma'} c^\dagger_{i\sigma} c^\pd_{j\sigma'}
\end{equation}
where $\Upsilon_{ij}=e^{i\Gamma_{ij}\frac{\phi_{ij}}{2}\sigma^z}$.
The in-plane SOC term $H_R$ contains a conventional Rashba splitting in addition to a radial Rashba effect,
\begin{equation}
\label{eqn:hr}
    H_R = H_{R, {\rm conv.}} + H_{R, {\rm rad}}
\end{equation}
    which can be more naturally expressed as
    \begin{equation}
    H_{R, {\rm conv.}} = \sum\limits_{i,j} \sum\limits_{\sigma,\sigma'}  [i\alpha_{ij}(\bm{\sigma}\times\bm{r}_{ij})\cdot\hat{\bm{z}} ]_{\sigma\sigma'}c^\dagger_{i\sigma} c^\pd_{j\sigma'}
\end{equation}
and
\begin{equation}
    H_{R, {\rm rad}} = \sum\limits_{i,j} \sum\limits_{\sigma,\sigma'}  [i\beta_{ij}\Gamma_{ij}(\bm{\sigma}\cdot\bm{r}_{ij})]_{\sigma\sigma'}c^\dagger_{i\sigma} c^\pd_{j\sigma'}
\end{equation}
where ${\alpha_{ij} = \lambda_{ij}\cos(\phi_{ij})}$, ${\beta_{ij} = \lambda_{ij}\sin(\phi_{ij})}$, and the form factor ${\Gamma_{ij} = -{\rm{sgn}}(\sin(6\theta_{ij}))}$. Here $\theta_{ij}$ is given by the position of each lattice site. The radial Rashba effect has been investigated in graphene-based heterostructures \cite{frankEmergenceRadialRashba2024b,kangMagnetotransportSignaturesRadial2024}, though not in triangular lattice systems. The strength of the Rashba splitting is given by $\lambda_{ij}$ and the mixing between conventional and radial Rashba terms is controlled by $\phi_{ij}$.

We only find finite $\phi$ mixings for further-nearest neighbor shells containing 12 sites, where the radial strengths $\beta_{ij}$ shown in Tab.\,\ref{table:Dot product splitting DFT+U} are comparable to the conventional $\alpha_{ij}$.
We also find what appears to be a valley Zeeman term, given by $H_{\rm{VZ}}$. This term is finite for nearest neighbors that do not lie on any of the mirror planes of the lattice. The sign of the term is governed by the form factor
$\nu_{ij} = -{\rm{sgn}}(\sin(3\theta_{ij}))$.
The strengths $J_{ij}$ given in Tab.\,\ref{table:DMI splitting DFT+U} show that this effect is quite weak, with $J_2/t_1$ less than 1\% in all materials.

\begin{figure}[t!]
    \centering
    \includegraphics[width=\linewidth]{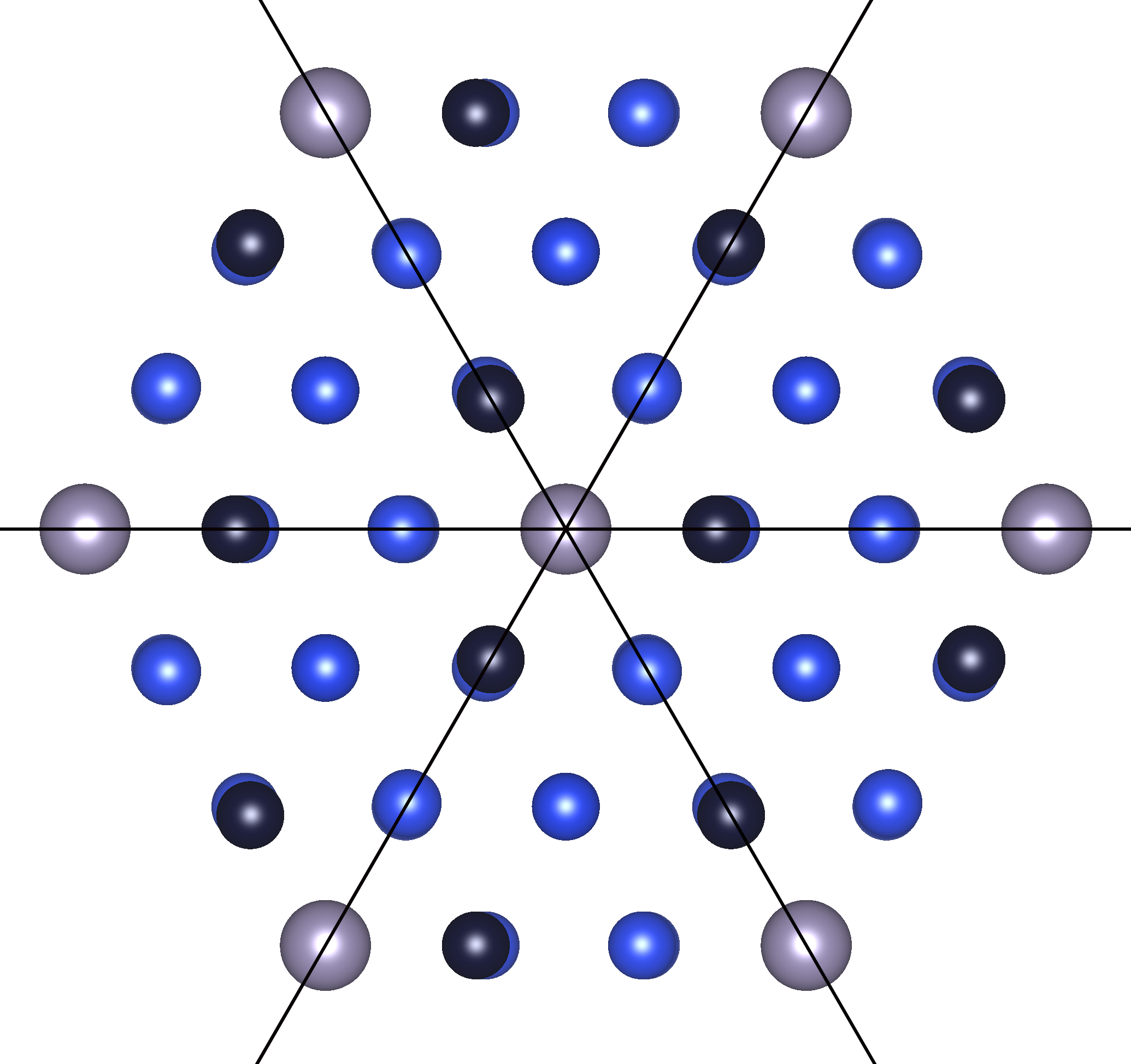}
    \caption{Top-down view of Sn/Si(111) with nearest neighbor Sn surface atoms (silver). Surface Si atoms (blue) have been darkened, emphasising how the bulk determines the choice of mirror planes (black lines) of the $C_{3v}$ symmetry.}
    \label{fig:C3v symmetry}
\end{figure}

\begin{figure}[t]
    \centering
    \includegraphics[width=\linewidth]{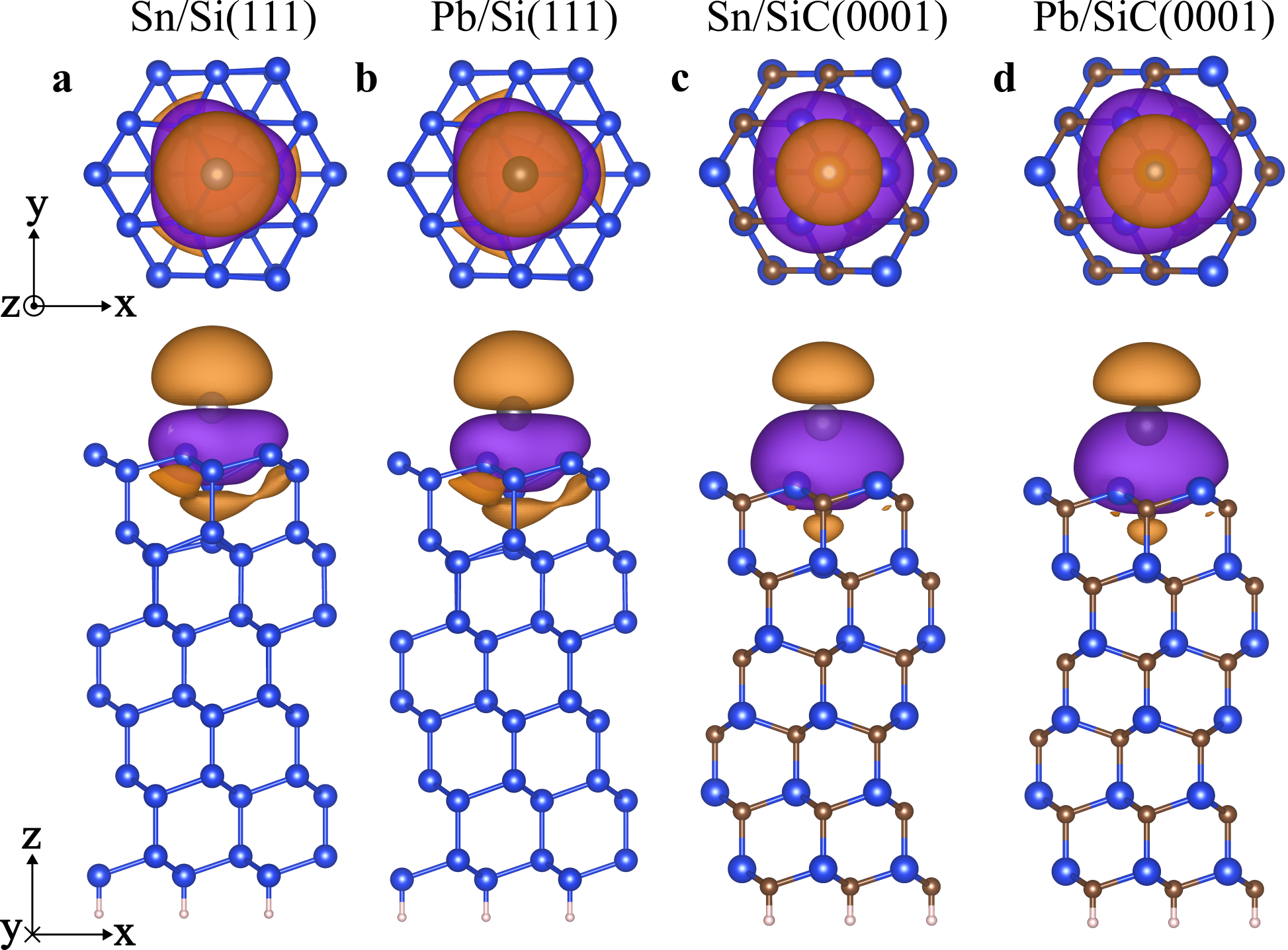}
    \caption{Maximally localized Wannier functions (MLWFs) for \textbf{a} Sn/Si(111), \textbf{b} Pb/Si(111), \textbf{c} Sn/SiC(0001) and \textbf{d} Pb/SiC(0001) are shown from top and side views. Adatoms are shown in grey, Sn in blue, C in brown, and H in pink. MLWFs are calculated using DFT+$U$ without the effects of SOC. We see low hybridization with the bulk in all materials, where the MLWFs only penetrate past the first layer.}
    \label{fig:wannier orbital mockup}
\end{figure}

Some of our derived tight-binding parameter deviate significantly from previously reported ones\,\cite{li-13nc1620,adler-19prl086401,badrtdinov-16prb224418,badrtdinovNanoskyrmionEngineeringElectron2018,glass-15prl247602,hansmann-13prl166401}. Most of the change is caused by the inclusion of $U$ in the DFT calculations. By increasing the gap size in order to match the experimentally observed values, the propagation of the Wannier functions is suppressed. The larger band gap acts as a wall resulting in highly localized Wannier functions and a decrease in hybridization with the bulk. Consequently our Wannier functions (shown in Fig.\,\ref{fig:wannier orbital mockup}\,a-d) are much more compact as compared with previous studies \cite{badrtdinov-16prb224418,glass-15prl247602}. This leads to a reduction in the longer-ranged hoppings, \eg\ $t_2/t_1$ from $-0.389$\,\cite{adler-19prl086401} to $-0.270$ for Sn/Si(111), corresponding to a 30\% decrease. While second and further neighbor hoppings are still non-negligible, these materials are closer to the nearest-neighbor triangular lattice than previously thought. More importantly, Rashba SOC can be quite sizeable for some of these materials. We note, however, that also our DFT calculations without inclusion of any $U$ led to some discrepancies with the literature. We assume that this is due to the higher resolution used in this work, as well as differing exchange-correlation functionals in the DFT calculations. In Appendix~\ref{sec:app-electronicstructure} we compare in Tab.\,\ref{tab:TB comparison} our derived parameters with previous works.

As previously mentioned, the presence of the substrate reduces the point group symmetry of the surface triangular lattice from $D_{6h}$ to $C_{3v}$, by breaking inversion symmetry and reducing the $C_6$ rotation symmetry to a $C_3$. This can be seen in Fig.\,\ref{fig:C3v symmetry}, which shows that the substrate also preserves the mirror planes which do not/do go through the bonds. We can further understand the terms in the Hamiltonian \eqref{eqn:ham} in three subsets by a hierarchy of symmetry breaking. The kinetic Hamiltonian $H_0$ preserves the full $D_{6h}$ point symmetry group. The Rashba SOC terms $H_R$ only break inversion symmetry and therefore have a $C_{6v}$ point group symmetry. Note that the radial Rashba effect is only $C_{6v}$ symmetric on 12 site coordination shells. The valley Zeeman-type terms $H_{\rm{VZ}}$ additionally break the the rotational symmetry and have a $C_{3v}$ symmetry. Anisotropic valley Zeeman terms symmetric under $C_{3v}$ are only possible on nearest-neighbor shells where the mirror planes do not intersect with bonds, and hence are only finite on these shells. We also note that additional Rashba coupling terms would be allowed under the reduction to a $C_{3v}$ point group symmetry, but these were not observed.

This symmetry hierarchy serves as one explanation of the relative strengths of the fitted parameters\,\cite{winkler2003}. Initially employed to understand bandstructures of bulk semiconductors under external magnetic field or uniaxial strain\,\cite{trebin1979, suzuki1974}, the symmetry hierarchy is the principle that terms in the Hamiltonian with lower symmetry typically have smaller prefactors than those with higher symmetry.
While this would generally mean that the Hamiltonian can be well approximated by the higher symmetry terms alone, the lower symmetry terms may still lead to important physical effects. In this case, the Rashba and the anisotropic valley Zeeman type terms are small, but follow Moriya's rules for antisymmetric anisotropic spin exchange\,\cite{moriya1960}. The Dzyaloshinskii–Moriya interaction\,\cite{moriya1960, moriya1960a, dzyaloshinsky1958} is an important factor for explaining weak anisotropic ferromagnetism in bulk materials, and is shown to drive a first order phase transition to a weakly ferromagnetic phase above a (second-order transition to) an antiferromagnetic phase. These types of interactions remain unexplored in the realm of surface states, and could have potential consequences for known spin-density wave phases in these materials\,\cite{li-13nc1620}.


\section{Results II: Phononic structure}\label{sec:results2}

\begin{figure*}
    \centering
    \includegraphics[width=\textwidth]{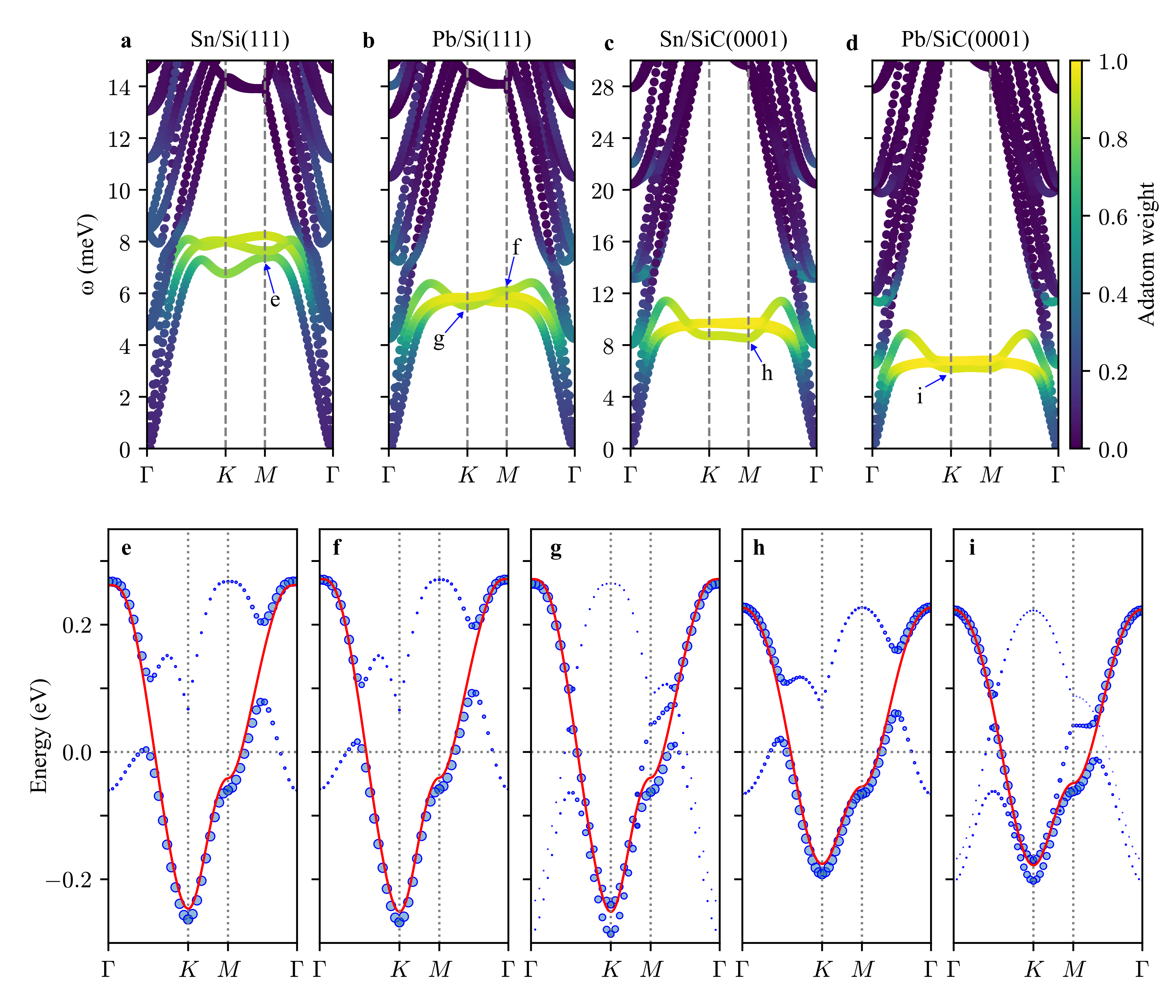}
    \caption{Phonon dispersions for \textbf{a} Sn/Si(111), \textbf{b} Pb/Si(111), \textbf{c} Sn/SiC(0001) and \textbf{d} Pb/SiC(0001). As with the electronic dispersions DFT+$U$ has been used for phonon calculations. The color of each point represents the adatom contribution to the mode (see color bar). Effective band structures for \textbf{e} Sn/Si(111), \textbf{f, g} Pb/Si(111), \textbf{h} Sn/SiC(0001) and \textbf{i} Pb/SiC(0001) for electron-phonon coupling calculations with 0.2\,\AA \,displacement amplitude. The selected phonon mode is indicated by the arrows in panels a-d. Electron-phonon coupling calculations have a reduced Brillouin zone due to the supercell required to encompass the effects of a given phonon mode. The band structure in the reduced Brillouin zone is thus unfolded onto the Brillouin zone of the original unit cell as an effective band structure. The band structure of the unperturbed unit cell is plotted in red for reference. The size of each blue dot represents band unfolding weight.}
    \label{fig:phonon_EPC}
\end{figure*}

Ultimately we are interested in the EPC strength; in order to compute it we need to derive the phonon spectra first.
For Sn/Si(111) that was previously done to explore the consequences for the correlated many-body phases\,\cite{zahedifar-19prb125427,wolf-22prl167002}.
Here we consistently ignore many-body phases such as magnetic or charge order, which would drastically affect the phonon bandstructures\,\cite{zahedifar-19prb125427}.
We use the software package phonopy \cite{phonopy-phono3py-JPCM,phonopy-phono3py-JPSJ} along with VASP to compute the phonon spectra, as shown in Fig.\,\ref{fig:phonon_EPC}\,a-d. We emphasize that the phonon computations are performed without spin-orbit coupling; the spin-orbit case is considered further below.
We have colored the phonon eigenvalues according to the adatom weight in the eigenmodes, so that the contribution of surface adatoms is emphasized.
Apart from the differences of frequencies $\omega$ at the $K$ and $M$ points, all four phonon spectra look essentially the same up to some global energy scaling. In particular, one can clearly see that the three Sn acoustic phonon bands disperse inside a gap into the Si(111) phonon dispersion. 
One should note the differing energy scales for X/Si(111) and X/SiC(0001) dispersions, showing higher energy SiC bulk modes \add{due to the lower average atomic mass in SiC compared to elemental Si}.

Computing these spectra turned out to be cumbersome. For X/Si(111) the phonon spectra can be straight-forwardly computed based on the DFT bandstructure using a $2\times 2$ supercell. For X/SiC(0001) we found, however, imaginary phonon modes. Relaxing the unit cell as well as switching from a $2\times 2$ to a $3\times 3$ supercell did not resolve the issue. We noticed that DFT underestimated the experimental gap of SiC(0001) by about 25\%. The previously mentioned DFT$+U$ treatment, where $U$ is applied to the atoms of the semiconductor substrate, allowed to adjust the band gap. By increasing the band gap and approaching the experimental value, the imaginary phonon modes shrink and disappear to give real modes when the experimental gap has been met (Fig.\,\ref{fig:phonon_EPC} c and d). As DFT underestimates the band gap of Si(111) by the same amount (25\%), we repeated the phonon calculations for X/Si(111) with DFT$+U$ input (shown in Fig.\,\ref{fig:phonon_EPC}\,a and b).

The phonon bands of the adatoms (\ie Sn and Pb) shown in Fig.\,\ref{fig:phonon_EPC}\,a-d have relatively flat dispersions, especially in the case of the X/SiC(0001) materials. There has been recent interest in flat phonon bands with regard to ferroelectrics\,\cite{leeScalefreeFerroelectricityInduced2020}. As the flat bands found here are energetically separated from the bulk bands, these materials could act as good platforms for further exploring flat phonon bands experimentally.


\section{Results III: Electron-Phonon Coupling}\label{sec:results3}

A fully first-principles study is computationally too expensive for the considered materials. Instead we perform an effective approach based on deformation potentials $\mathcal{D} = \Delta E_k/\Delta Q$ due to frozen phonon modes with mode amplitude $Q$. Loosely speaking, a phonon that strongly couples to states near the Fermi energy $E_F$ cause a large shift $\Delta E_k$ in the electronic bandstructure near $E_F$.

 Within the deformation potential approximation, the EPC of a specific phonon mode is calculated via\,\cite{khan-84prb3341} $\lambda = 2 N(E_F) \frac{\hbar}{2 M \omega^2}|\mathcal{D}|^2$
where $N(E_F)$ is the density-of-states per spin and unit cell, $M$ is the mass of the adatom, and $\mathcal{D}$ is the aforementioned deformation potential. By inspecting the equation for $\lambda$ one also notices that the lowest frequency modes contribute most to the EPC due to the $1/\omega^2$ suppression. Thus we only focus on the three lowest branches at the $K$ or $M$ high symmetry points. The smallest frequencies are $\omega = 5.52$meV, see Tab.\,\ref{tab:electron phonon coupling}. In the same table we also list the EPC strength $\lambda$ along with the values of the corresponding deformation potential as well as the DOS.

    For a given mode we first create a supercell large enough to realize the full size of the phonon. At the $M$ and $K$ modes this corresponds to $2\times2$ and $3\times3$ supercells containing 4 (160) and 9 (360) adatoms (atoms), respectively.
    The supercell atoms are then perturbed according to the chosen phonon mode eigenvector and the associated band structure is calculated. Due to the reduction in Brillouin zone size as we increase the size of our supercell, we must unfold the band structure onto the unit cell Brillouin zone to obtain an effective band structure. For each mode we perturb atomic positions by a maximum of $\Delta Q = 0.1 \rm{\AA}$ and $\Delta Q = 0.2\rm{\AA}$. It is important to choose only small displacement amplitudes to stay within the linear response regime. From the effective band structures we then calculate the deformation potentials by comparing the gap opening $2\Delta E_k$ with $\Delta Q$.
    In Fig.\,\ref{fig:phonon_EPC}\,e-i we show unfolded effective band structures with $\Delta Q = 0.2\rm{\AA}$ for different phonon modes and materials. For band structures coupled to modes with both phonon wavevector $q=M$ and $q=K$ we find distinct gap openings along the $\Gamma K$ and $M \Gamma$ paths. For each material and coupled mode we calculate deformation potentials for the different gap openings, the largest of these being used to calculate $\lambda$.
    Due to the large number of atoms required in the EPC calculations it was unfeasible for us to perform all these calculations including relativistic effects.


\begin{table}[h!]
    \centering
    \begin{tabular}{c c c c c c}
    \hline\hline
        \multirow{2}{*}{Material} & Mode  & \multirow{2}{*}{$\lambda$}  & \multirow{2}{*}{$\omega\, {\rm (meV)}$} & \multirow{2}{*}{$\mathcal{D}\, ({\rm eV}/\text{\AA})$ } & \multirow{2}{*}{$N(E_F)$} \\
         & (band $\#$) & & & \\
    \hline
     \multirow{2}{*}{Sn/Si(111)}  & \spbox{$M$}{$K$} (1) & 0.059 & 6.750 & 0.179 & \multirow{2}{*}{$2.39\frac{1}{\rm eV}$}\\

       & $M$ (1) & 0.135 & 7.391 & 0.297 \\

      \multirow{2}{*}{Pb/Si(111)} & \spbox{$M$}{$K$} (1)& 0.062 & 5.516 & 0.207 & \multirow{2}{*}{$2.19\frac{1}{\rm eV}$}\\

       & $M$ (3) & 0.069 & 6.100 & 0.241 \\

      \multirow{2}{*}{Sn/SiC(0001)} & \spbox{$M$}{$K$} (1) & 0.045 & 8.720 & 0.189 & \multirow{2}{*}{$2.71\frac{1}{\rm eV}$} \\

       & $M$ (1) & 0.074 & 8.537 & 0.238 \\

      \multirow{2}{*}{Pb/SiC(0001)} & \spbox{$M$}{$K$} (1) & 0.037 & 6.190 & 0.158 & \multirow{2}{*}{$2.82\frac{1}{\rm eV}$} \\

       & $M$ (1) & 0.073 & 6.217 & 0.224 \\
    \hline\hline
    \end{tabular}
    \caption{DFT$+U$ electron-phonon coupling parameters for the four materials: phonon mode and band index, EPC strength $\lambda$, frequency $\omega$ of the mode, deformation potential $\mathcal{D}$ and DOS at $E_F$.}
    \label{tab:electron phonon coupling}
\end{table}





\section{Relativistic computation of EPC}
\label{sec:results4}

Phonon calculations are much more costly than electronic bandstructure computations; performing them relativistically, \ie with spin-orbit coupling, is even more expensive. To restore the coherence of the paper, we consider at least for the material Pb/Si(111) the phonon band structure relativistically. Surprisingly, the phonon spectrum hardly changes and there are only minor quantitative changes. In particular, the deformation potential is slightly larger; however, also $\omega$ is somewhat larger. In combination, $\lambda$ is minimally increased as compared to the non-relativistic case. We find $\lambda=0.085$ at $M$ and $\lambda=0.059$ at $K$, compared to the values in Tab.\,\ref{tab:electron phonon coupling} only a mild change. Comparison of the phonon spectrum of Pb/Si(111) with and without  SOC is shown in Fig.\,\ref{fig:SM-phonon SOC}. Further discussion about the effect on $\lambda$ is given in Appendix~\ref{sec:app-EPC}, including numerical values in Tab.\,\ref{tab:SM-electron phonon coupling}. While the increase in $\lambda$ is insignificant, it is still somewhat counter-intuitive as one would have expected the EPC to shrink in the presence of SOC.


\section{Discussion}
\label{sec:discussion}

For all materials we find strongest coupling to phonon modes at the $M$ high symmetry point, due to the gap opening along the $M\Gamma$ path. We find that the electronic structure couples notably only to  phonon modes with a large out-of-plane adatom displacement. Modes with purely in-plane displacement have vanishing effects on the band structure. One such mode, the $M$ high symmetry point of the second branch in Sn/Si(111), produces almost no noticeable shift in the band structure and gives a deformation potential approximately 500 times smaller than the results listed in Tab.\,\ref{tab:electron phonon coupling}. Effects of DFT+$U$ and in-plane motion are discussed in the Appendix C in the SM along with relevant parameters for calculations in Tab.\,\ref{tab:SM-electron phonon coupling}.

Tab.\,\ref{tab:electron phonon coupling} clearly rules out strong EPC as known \eg\ from MgB$_2$\,\cite{an-01prl4366}. The X/SiC(0001) materials have such a small $\lambda$ that any phonon-mediated superconductivity is out of the picture. Only Sn/Si(111) with its phonon mode at the $M$ point has a somewhat larger EPC strength $\lambda=0.135$. To further analyze if such a value of $\lambda$ could possibly be responsible for the observed superconducting transition at $T_c=4.7$K, we try to estimate $T_c$ via the Allen-Dynes equation\cite{allenTransitionTemperatureStrongcoupled1975} $T_c = \langle\omega\rangle/1.20 \exp{(-\frac{1.04(1+\lambda)}{\lambda-\mu^\ast(1+0.62 \lambda)} )}$. Assuming $\omega \approx 100$K and a Coulomb pseudopotential in the typical range $\mu^\ast \sim 0.05 - 0.2$ \cite{carbottePropertiesBosonexchangeSuperconductors1990}, we find for Sn/Si(111) $T_c \sim 10^{-15}$K $\approx 0$. 

\begin{figure}[t!]
    \centering
    \includegraphics[width=\linewidth]{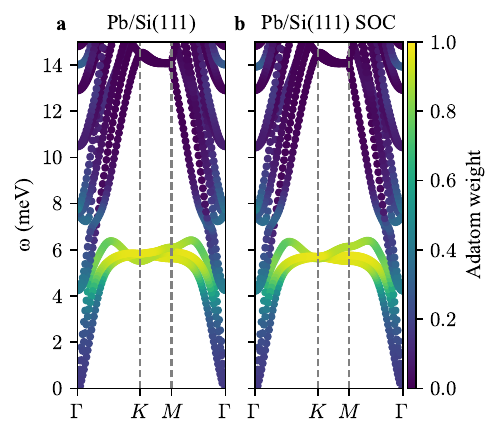}
    \setlength{\belowcaptionskip}{-20pt}
    \caption{Phonon dispersion for Pb/Si(111) \textbf{a} without and \textbf{b} with the SOC included. Both calculations were done using DFT+$U$.}
    \label{fig:SM-phonon SOC}
\end{figure}

It is worth emphasizing that the deformation potential method used here to estimate $\lambda$ replaces the momentum-dependence of EPC with its value at the respective high symmetry point, which is likely to overestimate EPC and thus overestimate $T_c$. Moreover, if we follow the ideas of Ref.\,\onlinecite{koch-99prl620} to quantify $\mu^\ast$ we would end up with an even larger value than $\mu^\ast \sim 0.05 - 0.2$, causing a further reduction of $T_c$. Despite the approximate character of estimating $T_c$ employed in this paper, we conclude that it does not appear to be plausible that EPC could be responsible for the observed superconductivity in Sn/Si(111)\,\cite{ming-18prb075403,ming-23np500}.

%
%

\section{Outlook}\label{sec:outlook}

We have analyzed the electronic and phononic properties of Sn/Si(111), Pb/Si(111), Sn/SiC(0001) and Pb/SiC(0001) with a $\sqrt{3}\times\sqrt{3}$ reconstructed surface phase. Using a deformation potential method, we find that the EPC strength $\lambda$ is too small in all materials to drive conventional superconductivity. We thus conclude that the recently experimentally observed superconductivity in Sn/Si(111) is likely to be driven by electron--electron interactions. This is in line with the recent experimental claim that the superconductivity is of chiral $d$-wave type. The tight-binding parameters derived in this paper will be the starting point for future quantum many-body computations of the filling-dependent interacting phase diagrams using methods such as random phase approximation, density matrix renormalization group or functional renormalization group.

\vspace{-0.5cm}
\section*{Data availability}
All data needed to evaluate the conclusions in the paper are
present in the paper and are available for download at Ref.\,\onlinecite{marchetti_2025_14957998}.
\vspace{-0.5cm}


\begin{acknowledgments}
We acknowledge discussions with Steve Johnston, Hanno Weitering and Hua Chen. 
S.R.\ acknowledges support from the Australian Research Council through Grant No.\ DP200101118 and DP240100168.
This research was undertaken using resources from the National Computational Infrastructure (NCI Australia), an NCRIS enabled capability supported by the Australian Government.
\end{acknowledgments}


\appendix
\section{Electronic structure}\label{sec:app-electronicstructure}

\begin{figure}[b]
    \centering
    \includegraphics[width=\linewidth]{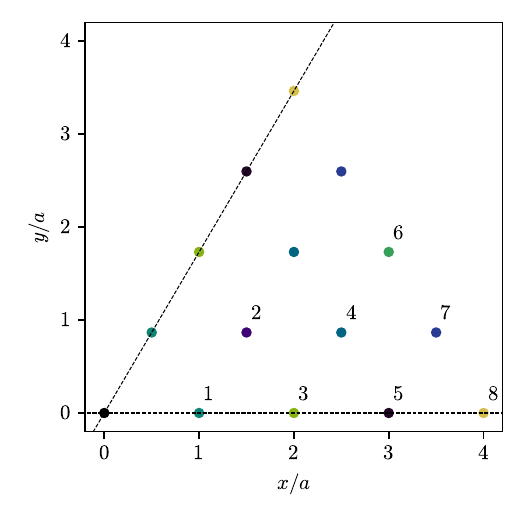}
    \caption{
        Schematic of the nearest-neighbors included in the surface state Wannierization. Symmetry planes preserved by the substrate are also shown ({\it cf}.\ Fig.\,\ref{fig:C3v symmetry}).
    }
    \label{fig:nearest neighbours}
\end{figure}
\begin{figure*}[t!]
    \centering
    \includegraphics{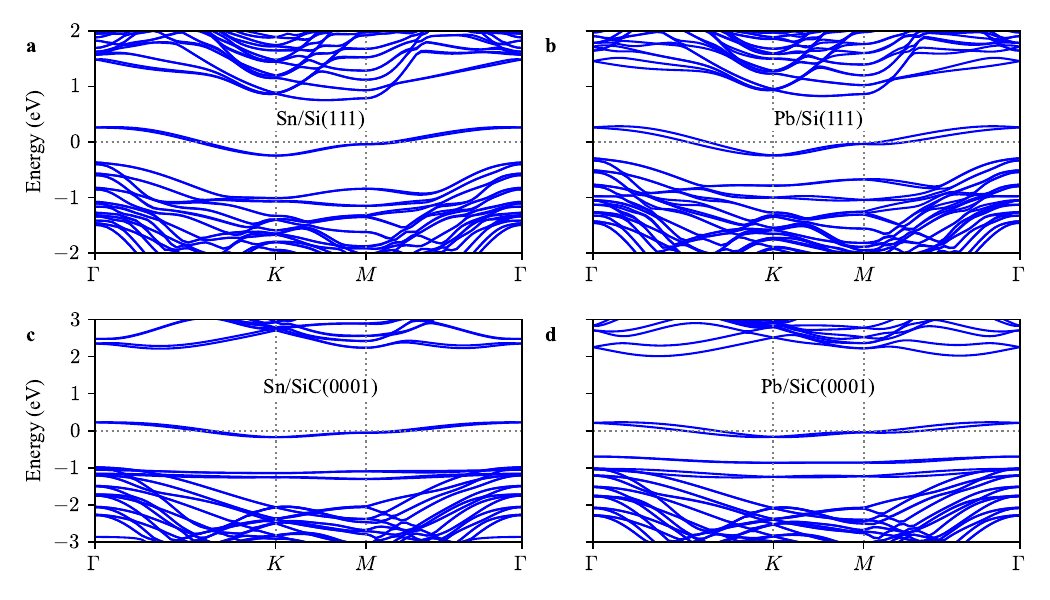}
    \caption{Same band structures as Fig.\,\ref{fig:DFT+U_TB} for \textbf{a} Sn/Si(111), \textbf{b} Pb/Si(111), \textbf{c} and Sn/SiC(0001) \textbf{d} Pb/SiC(0001) but with a wider energy range so that also substrate bands are visible.}
    \label{fig:SM-bandstructure}
\end{figure*}

\subsection{Group theory decomposition of Hamiltonian}\label{sec:appA}

In the main text of the manuscript, the symmetry hierarchy is discussed, whereby we can decompose terms of the Hamiltonian by successive symmetry reduction. In this section, we discuss practically how this is done, and go into further detail describing the possible symmetry allowed Hamiltonian terms.

A regular triangular Bravais lattice follow a $D_{6h}$ point group symmetry, which is generated by a $C_6$ rotation, inversion symmetry and a mirror plane $\sigma_1$. The presence of the substrate breaks inversion symmetry, as well as reduces the $C_6$ to a $C_3$ rotation symmetry. This can be seen in Fig.\,\ref{fig:C3v symmetry}, where the alternating heights of the substrates reduce this rotational symmetry, also removing half of the mirror planes.

All possible terms included in the Hamiltonian must be some linear combination of basis functions of the $A_1$ irrep of reduced symmetry point group $C_{3v}$. It is illuminating, however, to first construct the possible basis functions of irreps $C_{6v}$, which includes the spin-orbit coupling and inversion symmetry breaking. Then, we consider which $C_{6v}$ irreps map to $A_1$ in $C_{3v}$ upon the rotation symmetry reduction $C_6 \rightarrow C_3$. 

The Hamiltonian is written in terms of the bilinears $c^\dagger_{i \sigma} c^\pd_{j \sigma'}$, which represent an electron hopping from site $j$ with spin $\sigma'$ to site $i$ with spin $\sigma$. Translation invariance means that we need only consider the relative vector between the sites, \textit{i.e.}, we need only consider the symmetry group actions on $\bm{r}_{ij}$, where site $i$ is located at Bravais lattice point $\bm{R}_i$, and $j$ at $\bm{R}_j = \bm{R}_i + \bm{r}_{ij}$. Point group symmetry operations permute the bilinears amongst spin indices as well those with the same hopping distance $|\bm{r}_{ij}|$. We then only need to consider the nearest-neighbor shells with the same hopping distances one at a time, \textit{i.e.}, the relevant Hilbert space is $4N$, for $2 \times 2$ spin indices, and a nearest-neighbor shell with coordination number $N$.

The basis functions can be found by constructing the representation $R(g)$ of the symmetry group acting on this Hilbert space, forming projection operators
\begin{equation}
    P^\nu = \sum_g \frac{d_\nu}{|G|} \left( \chi^{\nu} (g) \right)^{*} R(g)
\end{equation}
for each irrep $\nu$, and diagonalizing to find the basis functions as the eigenvectors with magnitude 1. $d_\nu$ is the dimension of the irrep $\nu$, $|G|$ is the order of the group (which is 12 for $C_{6v}$), and $\chi^\nu (g)$ is the character of the group element $g$ in the irrep $\nu$.

The matrix representations of the symmetry elements 
can be constructed from a matrix outer product of the spatial transformation $R(g)$, which can be viewed as the transformation of the bond vector $\bm{r}_{ij}$, and the two spin transformations $S(g)$:
\begin{equation}
    R(g) \otimes S^{*} (g) \otimes S(g)
\end{equation}
where $c_{is} \rightarrow \sum_{s'} S_{ss'} c_{is'}$ is a spin transformation acting on a second quantized fermionic annihilation operator ($c^\dagger_{is} \rightarrow \sum_{s'} S^*_{ss'} c^\dagger_{is'}$ for the creation operator). $S(g)$ takes the usual form of the $SU(2)$ rotation operator $S(g) \equiv U(\vartheta, \hat{\bm{n}}) = \exp [-i \vartheta (\bm{\sigma} \cdot \hat{\bm{n}})/2]$.


We will now discuss the basis functions for the different nearest-neighbor shells. Fig.\,\ref{fig:nearest neighbours} shows the different nearest-neighbor shells, up to eighth nearest neighbors. We have included the set of mirror planes which are kept after the symmetry reduction $C_6 \rightarrow C_3$. From this figure, we can see that there are three distinct cases to consider. The first is the nearest-neighbors, which is a coordination shell of 6 where the bonds lie in mirror planes. The same analysis also applies to the third, fifth and eighth-nearest neighbor shells.

For the nearest-neighbors, the $C_{6v}$ symmetric Hamiltonian terms are basis functions of the $A_1$ irrep of $C_{6v}$:
\begin{align}
    \label{eqn:ht_sm}
    H_t &= -t \sum_{\langle ij \rangle, \, \sigma} c^\dagger_{i\sigma} c^\pd_{j\sigma}, \\
    \label{eqn:hj_sm}
    H_{J} &= - J \sum_{\langle ij \rangle, \, \sigma \sigma'} \sigma_{\sigma \sigma'}^z c^\dagger_{i\sigma}  c^\pd_{j\sigma'}, \\
    \label{eqn:hrconv_sm}
    H_{R, {\rm conv.}} &= - i \alpha \sum_{\langle ij \rangle, \, \sigma \sigma'}  [ (\bm{\sigma}\times\bm{r}_{ij})\cdot\hat{\bm{z}} ]_{\sigma \sigma'} \, c^\dagger_{i \sigma} c^\pd_{j \sigma'}
\end{align}
which we identify as the kinetic hopping, symmetric Zeeman splitting and the conventional Rashba terms respectively. Once we reduce the symmetry to $C_{3v}$, an additional term given by the $B_1$ irrep of $C_{6v}$ is allowed:
\begin{equation}
    H_{R, {\rm VR}} = i \beta \sum_{\langle ij \rangle, \, \sigma \sigma'}  \nu^\pd_{ij} \, (\bm{\sigma}\cdot\bm{r}_{ij})^\pd_{\sigma \sigma'} c^\dagger_{i\sigma} c^\pd_{j\sigma'}
\end{equation}
where $\nu_{ij} = - \text{sgn} ( \sin ( 3 \theta_{ij}))$ as defined in the main text, with $\theta_{ij}$ as the principle angle made by the bonding vector $\bm{r}_{ij}$. This term would correspond to a valley radial (VR) Rashba term.

The second case is the second nearest-neighbor shell, which also applies to the sixth nearest-neighbors. This is a shell with coordination number of six, however the bonds do not lie on the $C_{3v}$ mirror planes, but rather halfway between the planes. The $C_{6v}$ symmetric Hamiltonian terms are therefore the same, but the $C_{3v}$ symmetric term is given by the other $B$ irrep of $C_{6v}$, \ie $B_2$:
\begin{align}
    \label{eqn:hvt_sm}
    H_{\rm{Vt}} &= - t \sum_{\langle \langle ij \rangle \rangle, \, \sigma } \nu^\pd_{ij} \, c^\dagger_{i\sigma} c^\pd_{j\sigma} \\
    H_{\rm{VZ}} &= - J \sum_{\langle \langle ij \rangle \rangle, \, \sigma \sigma'} \nu^\pd_{ij} \, \sigma^z_{\sigma \sigma'} c^\dagger_{i\sigma} c^\pd_{j\sigma'} \\
    \label{eqn:hvrconv_sm}
    H_{R, {\rm conv.}} &= - i \alpha \sum_{\langle \langle ij \rangle \rangle, \, \sigma \sigma'} \nu^\pd_{ij} \, [ (\bm{\sigma}\times\bm{r}_{ij})\cdot\hat{\bm{z}} ]_{\sigma \sigma'} \, c^\dagger_{i \sigma} c^\pd_{j \sigma'}
\end{align}
which are a valley alternating versions of the kinetic hopping, the Zeeman term and the conventional Rashba term.

The third case is the forth nearest-neighbors, which also applies to the seventh and ninth nearest-neighbors. This is a shell with coordination number 12, where the mirror planes do not intersect any bonds. On top of the terms found for the 6 coordination number shells (Eq.\,\ref{eqn:ht_sm}-\ref{eqn:hrconv_sm}), we find that the following radial Rashba term is $C_{6v}$ symmetric:
\begin{equation}
    \label{eqn:hrrad_sm}
    H_{R, {\rm rad}} = i \beta \sum_{ \langle i,j \rangle_4, \, \sigma \sigma' }  \Gamma_{ij} (\bm{\sigma}\cdot\bm{r}_{ij})^\pd_{\sigma \sigma'} c^\dagger_{i\sigma} c^\pd_{j\sigma'}
\end{equation}
where ${\Gamma_{ij} = -\rm{sgn}(\sin(6\theta_{ij}))}$. When the symmetry is reduced to $C_{3v}$, the valley-alternating versions of the $C_{6v}$ symmetric terms are allowed. The forms of these were already given in Eq.\,\ref{eqn:hvt_sm}-\ref{eqn:hvrconv_sm}, and can be extended to the radial Rashba term Eq.\,\ref{eqn:hrrad_sm} by an analogous inclusion of the factor $\nu_{ij}$.

In all cases, all allowed $C_{6v}$ kinetic hopping and Rashba terms are observed, and only the symmetry reduced $C_{3v}$ valley Zeeman Hamiltonian term is observed, as discussed in the main text. This is then why the radial Rashba term is only observed on nearest-neighbor shells with coordination number 12, and why the valley Zeeman splitting is not observed on some shells, including nearest neighbors.

\begin{table*}[t!]
\centering
    \begin{tabular}{c| c c c c c c c c}
    \hline\hline
        Paper (method) & $t_1$ (meV) & $t_2/t_1$ & $t_3/t_1$ & $t_4/t_1$ & $t_5/t_1$ & $\alpha_1/t_1$ & $\alpha_2/t_1$ & $J_2/t_1$ \\
        \hline
        Sn/Si(111) & \\
        \textbf{This paper 2024 (PBE+U)} & $52.773$  & $-0.2703$ & $0.0974$ & $-0.0183$ & $0.0194$ & $0.1029$ & $0.0060$ & $0.0026$ \\
        \textbf{This paper 2024 (PBE)} & $54.917$  & $-0.3150$ & $0.1191$ & $-0.0236$ & $0.0210$ &  $0.1010$ & $0.0069$ & $0.0027$ \\
         F. Adler 2019\cite{adler-19prl086401} (LDA) & $52.2$ & $-0.3889$ & $0.1418$ & $-0.0268$ & & & & \\
         Li 2013\cite{li-13nc1620} (LDA) & $52.7$ & $-0.3881$ & $0.1555$ & $-0.0228$ & $-0.0318$ & & & \\
         D. I. Badrtdinov 2016\cite{badrtdinov-16prb224418} (PBE) & $43.51$ & $-0.4365$ & & & & $0.1271$ & $0.0198$ & $0.0032$ \\
         P. Hansmann 2013\cite{hansmann-13prl166401} (GW+DMFT) & $42.0$ & $-0.4762$ & $0.2381$ & & & & & \\
    \hline
        Pb/Si(111) & \\
        \textbf{This paper 2024 (PBE+U)} & $52.115$  & $-0.2764$ & $0.1005$ & $-0.0176$ & $0.0205$  & $0.2943$ & $0.0141$ & $0.0026$ \\
        \textbf{This paper 2024 (PBE)} & $54.09$  & $-0.3256$ & $0.1253$ & $-0.0234$ & $0.0227$  & $0.2907$ & $0.0153$ & $0.0011$ \\
         F. Adler 2019 (LDA) & $58.5$ & $-0.3829$ & $0.1248$ & $-0.0239$ & & & & \\
         D. I. Badrtdinov 2016 (PBE) & $41.32$ & $-0.4635$ & & & & $0.4037$ & $0.0506$ & $0.0027$ \\
         P. Hansmann 2013 (GW+DMFT) & $42.0$ & $-0.4762$ & $0.2381$ & & & & & \\
    \hline
        Sn/SiC(0001) & \\
        \textbf{This paper 2024 (PBE+U)} & $43.110$  & $-0.1983$ & $0.0199$ & $0.0026$ & $-0.0032$ & $0.1661$ & $0.0128$ & $0.0034$ \\
        \textbf{This paper 2024 (PBE)} & $39.767$  & $-0.2889$ & $0.0504$ & $0.0021$ & $-0.0070$  & $0.1767$ & $0.0195$ & $0.0024$ \\
        D. I. Badrtdinov 2018\cite{badrtdinovNanoskyrmionEngineeringElectron2018} (PBE) & $27.94$ & $-0.5140$ & & & & $0.2645$ & $0.0537$ & $0.0043$ \\
        S. Glass 2015\cite{glass-15prl247602} (LDA) & $27.3$ & & & & & & & \\
    \hline
        Pb/SiC(0001) & \\
        \textbf{This paper 2024 (PBE+U)} & $40.175 $ & $-0.2042$ & $0.0401 $& $-0.0005$ & $-0.0048$ &  $0.4440 $& $0.0318 $& $0.0072 $\\
        \textbf{This paper 2024 (PBE)} & $36.381 $ & $-0.3160$ & $0.0767 $& $-0.0026$ & $-0.0098$ &  $0.4901 $& $0.0501 $& $0.0009 $\\
    \hline\hline
         
    \end{tabular}
    \caption{Comparison of tight-binding Hamiltonian parameters with other publications.}
    \label{tab:TB comparison}
\end{table*}

\subsection{Full band structure}
In Fig.\,\ref{fig:SM-bandstructure} we show the same band structure as in Fig.\,\ref{fig:DFT+U_TB} in the main text but over a wider energy range. DFT+$U$ has been used to increase the semiconductor band gaps to match experiment, which is reflected in the figure. One can see that the surface bands disperse through the semiconductor band gap, and we note that there is no energy overlap between the bulk bands and the surface bands in any of the materials.
\add{
In the case of Pb/SiC(0001), another isolated band lies within the bulk gap at approximately $-0.75$\,eV. 
Orbital analysis of this band reveals it is of adatom origin, with strong Pb $p_x$ and $p_y$ character; hence, it was ignored for calculating the size of the bulk gap in Sec.\,\ref{sec:results1}.}

\subsection{Comparison of tight-binding parameters with previous works}\label{sec:appA3}

In this paper, we have extracted tight-binding parameters as computed with DFT but also with DFT+$U$ where $U$ was only applied to the substrate atoms. The resulting tight-binding parameter differ, and partly they do differ significantly. In Tab.\,\ref{tab:TB comparison} we compare our tight-binding parameters with previous results from the literature.

\section{Electron-phonon coupling}\label{sec:app-EPC}


In Tab.\,\ref{tab:SM-electron phonon coupling} we compare EPC calculations performed with and without DFT+$U$ for the X/Si(111) materials, in-plane phonon modes, and the effect of SOC in phonon calculations. We find that the inclusion of DFT+$U$ causes a reduction in the EPC strength $\lambda$. Notably, the deformation potential $\mathcal{D}$ is mostly unaffected, with the largest change $\Delta\mathcal{D} = 0.010\,$eV/\AA\, in Pb/Si(111) coupled to the band 1 $K$ mode. The source of this reduction can be mostly attributed to the general increase in phonon mode frequency seen in both materials for a given mode, as $\lambda \sim 1/\omega^2$.
For both Sn/Si(111) and Sn/SiC(0001) we test phonon modes with eigenvector corresponding to in-plane motion. For these calculations we find no discernible change in the effective band structure as compared to the unperturbed unit cell and thus minimal coupling. It is unsurprising that out-of-plane motion will have the largest effect on the band structures when considering hybridization of Wannier functions with the bulk. As discussed in the main text, the reduction in hybridization corresponds to significant changes in tight-binding hopping parameters. Changes in the adatom height above the substrate should have the largest effect on hybridization, while in-plane motion should leave this largely unchanged.

We also compare the effect of including SOC when calculating the phonon dispersion for Pb/Si(111). The phonon modes corresponding to out-of-plane motion have minor changes in frequency. We see a slight increase in EPC strength $\lambda$ when coupled to the band 1 $M$ mode, which is due to the increase in deformation potential $\mathcal{D}$. As changes in $\lambda$ are minor in Pb/Si(111) which has a large Rashba splitting $\alpha_1/t_1 \sim 30\%$, we suspect that including SOC in the phonon calculations will have minimal effects on the other materials.

\begin{table}[t!]
    \centering
    \begin{tabular}{c c c c c }
    \hline\hline
        \multirow{2}{*}{Material} & Mode  & \multirow{2}{*}{$\lambda$}  & \multirow{2}{*}{$\omega\, {\rm (meV)}$} & \multirow{2}{*}{$\mathcal{D}\, ({\rm eV}/\text{\AA})$ }\\
         & (band $\#$) & & \\
    \hline
        Sn/Si(111) & K (1) & 0.091 & 5.244 & 0.175 \\

        Sn/Si(111) DFT+$U$ & K (1) & 0.059 & 6.750 & 0.179 \\

        Sn/Si(111) & M (1) & 0.189 & 6.196 & 0.297 \\

        Sn/Si(111) DFT+$U$ & M (1) & 0.135 & 7.391 & 0.297 \\

        Sn/Si(111) DFT+$U$ & M (2) & - & 6.995 & $\approx0$ \\
    \hline
        Pb/Si(111) & K (1) & 0.124 & 3.920 & 0.197 \\

        Pb/Si(111) DFT+$U$ & K (1) & 0.062 & 5.516 & 0.207 \\

        Pb/Si(111) DFT+$U$ & M (3) & 0.069 & 6.100 & 0.241 \\

        Pb/Si(111) SOC-ph & K (1) & 0.059 & 5.628 & 0.202 \\

        Pb/Si(111) SOC-ph & M (1) & 0.085 & 6.139 & 0.270 \\
    \hline
        Sn/SiC(0001) DFT+$U$ & K (1) & 0.045 & 8.720 & 0.189 \\

        Sn/SiC(0001) DFT+$U$ & M (1) & 0.074 & 8.537 & 0.238 \\

        Sn/SiC(0001) DFT+$U$ & K (2) & - & 9.683 & $\approx0$ \\
    \hline
        Pb/SiC(0001) DFT+$U$ & K (1) & 0.037 & 6.190 & 0.158 \\

        Pb/SiC(0001) DFT+$U$ & M (1)  & 0.073 & 6.217 & 0.224 \\
    \hline\hline
    \end{tabular}
    \caption{Full table of electron phonon coupling parameters. Calculations performed with and without DFT+$U$ are shown for X/Si(111). For Pb/Si(111) we also compare values obtained when including the SOC in phonon calculations. These calculations were also done with DFT+$U$. Calculations for modes corresponding to in-plane adatom displacement are also shown, with no notable deformation potential $\mathcal{D}$ observed.}
    \label{tab:SM-electron phonon coupling}
\end{table}

\begin{figure*}[t!]
    \centering
    \includegraphics[scale=1]{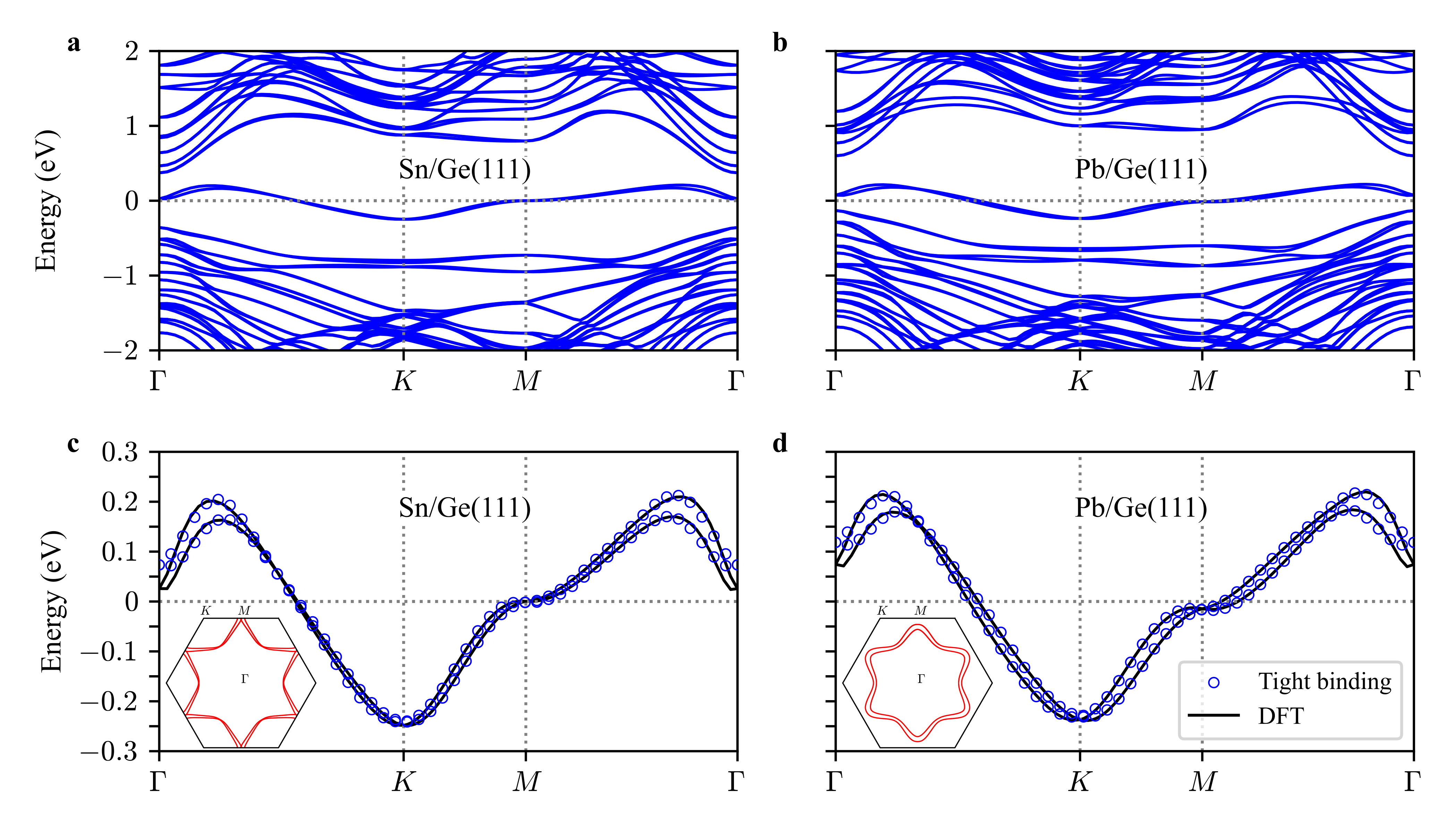}
    \caption{\add{DFT+$U$ bandstructures for \textbf{a} Sn/Ge(111) and \textbf{b} Pb/Ge(111). A zoomed in view of the surface bands along with tight-binding models are shown for \textbf{c} Sn/Ge(111) and \textbf{d} Pb/Ge(111). Up to 10th nearest neighbor terms have been used for the tight-binding models. Adding further nearest neighbor terms gives better matches around the $\Gamma$-point. Fermi surfaces have been included as insets, with the Fermi surface of Sn/Ge(111) sitting very close to the van Hove singularity.}}
    \label{fig:SM-Ge bands}
\end{figure*}

\section{Germanium substrate}\label{sec:App-Ge}
\begin{table*}[!t]
\centering
\addtolength{\tabcolsep}{-0.14em}
\begin{tabular}{c L L  c L L  c L L  c L L}
 \hline\hline
  \multicolumn{1}{c}{Hopping} & \multicolumn{1}{c}{Sn/Ge(111)} & \multicolumn{1}{c}{Pb/Ge(111)} & \multicolumn{1}{c}{SOC} & \multicolumn{1}{c}{Sn/Ge(111)} & \multicolumn{1}{c}{Pb/Ge(111)} & \multicolumn{1}{c}{SOC} & \multicolumn{1}{c}{Sn/Ge(111)} & \multicolumn{1}{c}{Pb/Ge(111)} & \multicolumn{1}{c}{SOC} & \multicolumn{1}{c}{Sn/Ge(111)} & \multicolumn{1}{c}{Pb/Ge(111)} \\
 \hline
    $t_0$ (eV) & 0.241074  & 0.122352 & & & & & & & & &  \\

    $t_1$ (eV)& 0.041324 & 0.043321 & $\alpha_1 / t_1$ & 0.019537 & 0.067036 & $\beta_1 / t_1$ & & & $J_1/t_1$  \\

    $t_2/t_1$ & -0.484682 & -0.400106 & $\alpha_2 / t_1$ & -0.050995 & 0.064605 & $\beta_2 / t_1$ & & & $J_2/t_1$ & 0.042881 & 0.042658 \\

    $t_3/t_1$ & 0.138612 & 0.118972 & $\alpha_3 / t_1$ & 0.001301  & 0.000853 & $\beta_3 / t_1$ & & & $J_3/t_1$ \\

    $t_4/t_1$ &  -0.094473 & -0.073313 & $\alpha_4 / t_1$ & 0.009354  & -0.010424 & $\beta_4 / t_1$ & -0.000007 & 0.003841 & $J_4/t_1$ & 0.008058 & 0.006787 \\

    $t_5/t_1$ & -0.013309 & -0.002355 & $\alpha_5 / t_1$ & 0.002634  & -0.002310 & $\beta_5 / t_1$ & & & $J_5/t_1$   \\

    $t_6/t_1$ & -0.090335 & -0.078438 & $\alpha_6 / t_1$ & -0.005791  & 0.007177 & $\beta_6 / t_1$ & & & $J_6/t_1$ & 0.007211 & 0.006879 \\

    $t_7/t_1$ & -0.016891 & -0.009418 & $\alpha_7 / t_1$ & 0.001802  & -0.001762 &  $\beta_7 / t_1$ & 0.000073 & -0.000223 & $J_7/t_1$ & 0.001162 & 0.001339 \\

    $t_8/t_1$ & -0.018948 & -0.015835 & $\alpha_8 / t_1$ & 0.001201  & 0.001412 & $\beta_8 / t_1$ & & & $J_8/t_1$ \\

    $t_9/t_1$ & -0.025869 & -0.021675 & $\alpha_9 / t_1$ & 0.001628  & -0.001869 &  $\beta_9 / t_1$ & -0.000016 & 0.000308 & $J_9/t_1$ & 0.001960 & 0.001962  \\

    $t_{10}/t_1$ & -0.013261 & -0.010480 & $\alpha_{10} / t_1$ & 0.000765  & -0.000824  & $\beta_{10} / t_1$ & -0.000045 & 0.000072 & $J_{10}/t_1$ & 0.000121 & -0.000023 \\
\hline\hline

\end{tabular}
 \caption{\add{Tight-binding parameters for X/Ge(111). Up to 10th nearest neighbor terms are listed for the hopping, conventional Rashba, radial Rashba, and valley-Zeeman parameters.}}
 \label{table:SM-Ge hoppings wide}
\end{table*}

For completeness we performed the same calculations in the main text to the materials Sn/Ge(111) and Pb/Ge(111). First-principles calculations for these materials can be found in the literature, see for instance Refs.\,\onlinecite{carpinelli-96n398,carpinelli-97prl2859,schuwalow-10prb035116,tresca-21prb045126,profeta-07prl086401}.

The use of DFT+$U$ is essential for X/Ge(111) as germanium is predicted as metallic in conventional DFT.
We use values of $U_{\rm Sn/Ge} = -7.25\,{\rm eV}$ and $U_{\rm Pb/Ge} = -7.86\,{\rm eV}$ to match germanium's $\Gamma$-point gap.

The structures were relaxed following the method for X/SiC(0001) outlined in the main text, with $10 \,{\rm \AA}$ of vacuum used for both materials. 
We find that the adatoms sit above the substrate at heights of $h = 1.95\,{\rm \AA}$ for Sn/Ge(111) and $h = 1.87\,{\rm \AA}$ for Pb/Ge(111).

The surface bands are shown in Fig.\,\ref{fig:SM-Ge bands} and have slightly larger bandwidths than the materials in the main text with $W/t_1 = 11.10$ $(W=0.46\,\rm{eV})$ for Sn/Ge(111) and $W/t_1 = 10.58$ $(W=0.46\,\rm{eV})$ for Pb/Ge(111).

The surface bands in both X/Ge(111) material show a dip around the $\Gamma$-point.
A denser $16\times 16 \times 1$ $\Gamma$-centered grid was required to capture this feature during the Wannierization process. 
Tight-binding parameters including all SOC terms are listed in Tab.\,\ref{table:SM-Ge hoppings wide}.
The tight-binding models plotted in Fig.\,\ref{fig:SM-Ge bands}\,c,d have up to 10th nearest neighbor contributions as listed in Tab.\,\ref{table:SM-Ge hoppings wide}.
This gives a good fit around the $K$ and $M$ high-symmetry points, and adding further nearest neighbor terms gives increasingly better fits around the $\Gamma$-point.
Rashba SOC is weaker compared to the materials discussed in the main text, with $\alpha_1/t_1=0.02$ for Sn/Ge(111) and $\alpha_1/t_1=0.07$ for Pb/Ge(111).
Both materials feature quite large $t_2/t_1$ and $\alpha_2/t_1$ values (compared to the other materials), leading to the dip around the $\Gamma$-point and another crossing of the bands along the $\Gamma K$ path.
Compared to X/Si(111) and X/SiC(0001) the van Hove singularities sit much closer to the Fermi surface, especially in the case of Sn/Ge(111).

\begin{figure}[b]
    \centering
    \includegraphics[width=\linewidth]{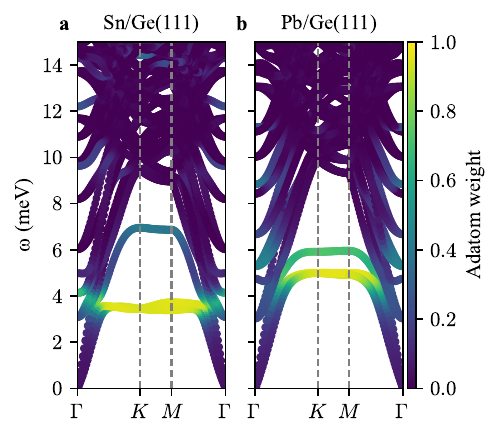}
    \caption{\add{Phonon dispersions for \textbf{a} Sn/Ge(111) and \textbf{b} Pb/Ge(111). The band corresponding to out-of-plane motion of the adatom sits above the two in-plane bands. Bulk modes sit at lower frequencies compared to X/Si(111) and X/SiC(0001) due to the higher atomic mass of Ge.}}
    \label{fig:SM-Ge phonons}
\end{figure}

Phonon dispersions are calculated using $2\times2$ supercells and DFT+$U$ and are displayed in Fig.\,\ref{fig:SM-Ge phonons}. 
In contrast to X/Si(111) and X/SiC(0001), the X/Ge(111) materials have out-of-plane surface bands sitting at higher frequencies than the in-plane modes.
In addition, the more massive germanium atoms lead to the bulk modes appearing at lower frequencies.

We calculate electron phonon coupling strengths for these materials following the same method outlined in the main text, with results listed in Tab.\,\ref{tab:SM-Ge EPC}.
We focus on the out-of-plane phonon modes and find that the EPC strengths $\lambda$ are too low for conventional superconductivity.
The density of states at the Fermi energy is comparatively much higher in the germanium materials as the van Hove singularity lies very close to the Fermi level, though this is balanced by the low deformation potentials $\mathcal{D}$.

\begin{table}[!h]
    \centering
    \begin{tabular}{c c c c c c}
    \hline\hline
        \multirow{2}{*}{Material} & Mode  & \multirow{2}{*}{$\lambda$}  & \multirow{2}{*}{$\omega\, {\rm (meV)}$} & \multirow{2}{*}{$\mathcal{D}\, ({\rm eV}/\text{\AA})$ } & \multirow{2}{*}{$N(E_F)$} \\
         & (band $\#$) & & & \\
    \hline
     \multirow{2}{*}{Sn/Ge(111)}  & \spbox{$M$}{$K$} (3) & 0.013 & 6.957 & 0.057 & \multirow{2}{*}{$5.57\frac{1}{\rm eV}$}\\

       & $M$ (3) & 0.046 & 6.873 & 0.106 \\

      \multirow{2}{*}{Pb/Ge(111)}  & \spbox{$M$}{$K$} (3) & 0.025 & 5.920 & 0.111 & \multirow{2}{*}{$3.63\frac{1}{\rm eV}$}\\

       & $M$ (3) & 0.051 & 5.962 & 0.158 \\

    \hline\hline
    \end{tabular}
    \caption{\add{Electron-phonon coupling parameters for X/Ge(111): phonon mode and band index, EPC strength $\lambda$, frequency $\omega$ of the mode, deformation potential $\mathcal{D}$ and DOS at $E_F$.}}
    \label{tab:SM-Ge EPC}
\end{table}

\clearpage

\bibliography{references}

\end{document}